\documentclass[12pt]{article}

\usepackage{psfrag}
\usepackage{amsfonts,amsmath}
\usepackage{latexsym}
\usepackage{amsthm}
\usepackage{amscd}
\usepackage{epsfig}
\newtheorem{theorem}{Theorem}[section]

\newtheorem{lemma}[theorem]{Lemma}

\newcommand\Diff{\operatorname{Diff}}
\newcommand\Homeo{\operatorname{Homeo}}

\newcommand\ohd{\frac{1}{2d}}

\newcommand{\rf}[1]{(\ref{#1})}
\newcommand{\oh}{\frac{1}{2}}

\newcommand{\beq}{\begin{equation}}
\newcommand{\eeq}{\end{equation}}
\newcommand{\bea}{\begin{eqnarray}}
\newcommand{\eea}{\end{eqnarray}}
\newcommand{\beas}{\begin{eqnarray*}}
\newcommand{\eeas}{\end{eqnarray*}}
\newcommand{\beqs}{\begin{displaymath}}
\newcommand{\eeqs}{\end{displaymath}}

\newcommand{\ep}{\varepsilon}

\newcommand{\ben}{\begin{equation}}
\newcommand{\een}{\end{equation}}

\newcommand{\bdm}{\begin{displaymath}}
\newcommand{\edm}{\end{displaymath}}

\newcommand{\pa}{\partial}
\begin{document}
 \headheight 0pt
 \topskip 0mm

 \addtolength{\baselineskip}{0.4\baselineskip}

 \pagestyle{empty}

 \vspace{0.5cm}

\hfill RH-05-00

\hfill October 2000

\vspace{1cm}

\begin{center}

{\Large \bf Discrete approximations to }

\medskip

{\Large\bf integrals over unparametrized paths}

%\vspace{1 truecm}

%\centerline{}

\vspace{1 truecm}

%\vspace{1.5cm}

\vspace{1.2 truecm}

Bergfinnur Durhuus\footnote{e-mail: durhuus@math.ku.dk}

Matematisk Institut

Universitetsparken 5, 2100 Copenhagen \O

Denmark

 \vspace{1.2 truecm}

  Thordur Jonsson\footnote{e-mail: thjons@raunvis.hi.is} 

 University of Iceland

 Dunhaga 3, 107 Reykjavik

 Iceland

 \vspace{1.2 truecm}

 \end{center}

 \noindent
 {\bf Abstract.} 
We discuss measures on spaces of unparametrized paths
related to the Wiener measure.  These measures arise naturally in
the study of
one-dimensional gravity coupled to scalar fields. Two kinds of discrete
approximations are defined, the piecewise linear and the hypercubic
approximations.  The convergence of these approximations 
in the sense of weak convergence of
measures is proven.  
%We discuss discrete approximations to the Wiener measure and related measures
%on unparametrized paths and show that the approximations
%converge weakly.  
We describe a family of sets of unparametrized paths that are analogous
to cylinder sets of parametrized 
paths. Integrals over some of these sets are evaluated in terms of 
Dirichlet propagators
in bounded regions.

 \vfill

 \newpage

 \pagestyle{plain}

 \section{Introduction}
In quantum field theory and string theory one frequently encounters the 
problem of integrating over geometrical objects, e.g., Riemannian manifolds
or Riemannian manifolds with some additional structure.
One wishes to define a measure on sets of geometrical objects and 
integrate functions that are independent of the coordinates used to describe 
the objects.  The prime example of a theory where this problem arises 
is the path integral
quantization of general relativity where one attempts to give meaning to
expressions of the form
\begin{equation*}
\langle F\rangle= \int  e^{-S(g)} F([g]) D[g],
\end{equation*}
where $g$ is a Riemannian metric on a manifold $M$, $[g]$ is the 
equivalence class of $g$ under diffeomorhisms of $M$, $F$ is a 
function and $S(g)$ is a 
diffeomorphism invariant action functional, e.g., the Einstein--Hilbert 
action \cite{dewitt,hawking}.  
Giving a mathematical meaning to expressions of this form is
largely an unsolved problem but some headway has been made, mainly 
in two dimensions, see \cite{book} and references therein.

One of the strategies used in physics to deal with functional integrals
of this type is to introduce discretizations of the geometrical objects
under consideration 
and try to prove convergence of the discretization as a cutoff parameter,
e.g., a lattice spacing, is taken to zero.  It inspires confidence in the results
obtained when different discretizations 
lead to identical continuum results.  This approach is described in detail
in the monograph \cite{book}.  

For one-dimensional objects, 
i.e., when the functional integral is over paths, the
situation is radically different from the higher dimensional analogues, since
we have measures on parametrized paths in $\mathbb R^d$ (for example 
the Wiener
measure) that are mathematically well-understood and give rise to measures on
unparametrized paths as we shall discuss below.
We study two different discretizations of integrals over 
unparametrized paths and show that the discrete measures converge to
the appropriate continuum measure.  

In ordinary quantum field theory applications  
of random paths it is often the convergence of regularized propagators that is
of  main interest and various results in this vein have been known for a long
time.  Our main interest is the 
convergence of the underlying measures on unparametrized paths, whereas
convergence of propagators merely means convergence of the total volume of the
measures.
%which has to our knowledge not been discussed before.
Corresponding problems in
non-relativistic quantum mechanics  normally involve only parametrized
paths.  In this case various aspects of discrete approximations pertaining
to the Wiener-measure on paths parametrized by a finite time interval have
been discussed by many authors, see, e.g., \cite{varadarajan} 
and references therein.

This paper is organized as follows.  In the next section we introduce the
models of discretized random paths we wish to study and give a proof of
pointwise convergence of the lattice propagator to the continuum propagator,
that will be needed later.  In section 3 we define the
appropriate path spaces, the continuum measures  and the discretized measures.
In section 4 we use  
standard tools of probability theory to prove the convergence of the 
discretized measures.  In section 5 we determine a family of sets of
unparamterized paths that generates the Borel sets of unparametrized paths and 
plays a role similar to the one played by cylinder sets for the Wiener-measure.
Finally, in section 6 we apply the results of the previous sections to
evaluate the measure of some of these sets.

\section{Propagators}

Let $\Delta$ denote the Laplacian in $\mathbb R^d$.  It is well-known that
the Euclidean propagator
\begin{equation}\label{2}
G(x,y)=2(-\Delta + m^2)^{-1}(x,y)
\end{equation}
of a scalar particle of mass $m>0$ in $\mathbb R^d$ has the path integral
representation
\begin{equation}\label{3}
G(x,y)=\int_{\omega:x\to y} e^{-m |\omega |}\,D\omega ,
\end{equation}
where $\omega$ is a path from $x$ to $y$ in $\mathbb R^d$ and $|\omega |$ 
denotes its length.  The most straightforward interpretation of the formal 
expression on the right hand side of  Eq.\ 
(\ref{3}) is obtained by regarding it as
a limit of lattice propagators.  We replace
$\mathbb R^d$ by the hypercubic lattice $a\mathbb Z^d$ with lattice spacing
$a$ and define a lattice propagator as
\begin{equation}\label{4}
 G^a(x,y)=a^{2-d}\sum_{\omega :x\to y}e^{-m(a)|\omega |}
\end{equation}
for $x,y\in a\mathbb Z^d$ where the sum is over all 
lattice paths from $x$ to $y$.
%We let $ax_a\to x$ and $ay_a\to y$ as $a\to 0$.
The prefactor $a^{2-d}$ is dictated by dimensional considerations and 
the dependence of the parameter $m(a)$ (lattice mass) on $a$ is determined 
by the requirement that $G^a(x,y)$ converge to $G(x,y)$ as $a\to 0$.

Using translation invariance we may set $G(x,y)=G(x-y)$ and
$ G^a(x,y)= G^a(x-y)$.  
The Fourier transform of the lattice propagator is then
\begin{eqnarray}
\widehat{ G^a}(k) & = & a^d\sum_{x\in a\mathbb Z^d}G^a(x)e^{-ik\cdot x}\nonumber\\
  & = & e^{am(a)}\left( m^2 +2a^{-2}\sum_{j=1}^d(1-\cos (ak_j))\right)^{-1}
\end{eqnarray}
where $k\in [-\pi /a,\pi /a]^d$ and $m(a)$ is given by the equation
\begin{equation}\label{5}
e^{am(a)}=2d+m^2a^2.
\end{equation}
Evidently this implies the desired uniform convergence in momentum space
\begin{equation}\label{6}
d^{-1}\widehat{ G^a}(k)\to 2(k^2+m^2)^{-1}= \widehat{G}(k)
\end{equation}
as $a\to 0$, for any $k\in \mathbb R^d$.  
%It is likewise easy to prove convergence
%in operator norm, see e.g.\ \cite{}.

Pointwise convergence in space-time can be obtained as follows.  We extend 
the lattice propagator from $a\mathbb Z^d$ 
to a smooth function on $\mathbb R^d$ by setting
\begin{equation}\label{7}
G^a(x)=\frac{1}{ (2\pi )^d}\int _{[-\pi /a,\pi /a]^d}\widehat{ G^a}(k)
e^{-ik\cdot x}\, dk
\end{equation}
for any $x\in \mathbb R^d$.
  For $\alpha =(\alpha _1,\ldots ,\alpha _d)$, where the
$\alpha _i$'s are non-negative integers, let
\begin{equation*}
\partial^\alpha=\prod_{i=1}^d\frac{\partial^{\alpha_i}}{\partial x_i^{\alpha _i}}.
\end{equation*}
Defining $\widehat{ G^a}(k)=0$ outside $[-\pi /a,\pi /a]^d$ it is easily
verified that
\begin{equation*}
\partial^\alpha \widehat{ G^a}(k)\to \partial^\alpha \widehat{G}(k)
\end{equation*}
uniformly on $\mathbb R^d$ for any multiindex $\alpha$.
Moreover, there is a constant $c_\alpha$ such that
\begin{equation}\label{int}
|\partial^\alpha 
\widehat{ G^a}(k)|\leq c_\alpha (k^2+m^2)^{-1-|\alpha |/2}
\end{equation}
where $|\alpha |=\alpha _1+\ldots +\alpha _d$.  Thus, choosing $|\alpha |>d$,
the right hand side of Eq.\ \rf{int} is integrable so 
the dominated convergence theorem together with Fourier inversion implies that
\begin{equation}\label{11}
d^{-1}x^\alpha  G^a(x)\to x^\alpha G(x)
\end{equation}
as $a\to 0$, where $x^\alpha =x_1^{\alpha _1}\ldots x_d^{\alpha _d}$.
In particular,
\begin{equation}\label{12}
d^{-1} G^a(x)\to G(x)
\end{equation}
for $x\neq 0$.

There is another path integral representation of the propagator $G(x,y)$
introduced in \cite{divecchia} 
whose analogue for surfaces has played an important
role in string theory in recent years \cite{polyakov}.  The alternative 
representation is given by
\begin{equation}\label{13}
G(x,y)=\int_{\omega : x\to y}\exp\left(-\frac{1}{2}\int \biggl(|\dot{\omega
    }|^2e^{-1} + m^2e\biggr)\,dt\right)\,D\omega De,
\end{equation}
where the integration is over paths $\omega$ in $\mathbb R^d$ from $x$ to $y$
and over intrinsic metrics $e$ on the paths. An intrinsic metric on the 
path is simply a positive definite function defined on the path.  
In order to give a
meaning to Eq.\ \rf{13}, we note an important common feature
of the two action functionals
\begin{equation}\label{14}
S_1(\omega) = m|\omega| = m\int|\dot{\omega}|\,dt
\end{equation}
and
\begin{equation}\label{15}
S_2(\omega,e) = \oh\int\biggl(|\dot{\omega}|^2 e^{-1}+m^2e\biggr)\,dt
\end{equation}
which occur in the path integrals \eqref{3} and  \eqref{13}.
The actions are invariant
under reparametrizations 
\begin{eqnarray}
 t' & = &\varphi(t)\nonumber\\
\omega'(t')& = &\omega(t)\nonumber\\
 e'(t')& = &\frac{e(t)}{\dot{\varphi}(t)}\,,\label{16}
\end{eqnarray}
where $\varphi$ is an increasing diffeomorphism between intervals. Thus the
path integrations in  Eqs.\ \eqref{3} 
and  \eqref{13} should be regarded as being
taken over diffeomorphism classes of paths in the first case and over 
diffeomorphism classes of paths and
metrics in the second one. 
The standard method for dealing with functional integration over such 
orbit spaces is the so called Faddeev-Popov procedure.  We
discuss the orbit spaces and the appropriate measures on them 
more thoroughly in
Section 3. For the moment we note that any pair $(\omega,e)$ can uniquely be
reparametrized to $(\omega',e')$ such that the parameter interval of the
latter is $[0,1]$ and the metric $e'$ is constant on $[0,1]$ and
equal to the volume
\begin{equation}
T\equiv \int e(t)dt
\end{equation}
of $e$, which is parametrization independent. It follows that the path
integral \eqref{13} is effectively an integral over $T$ and over paths
$\omega$ parametrized on $[0,1]$. An interpretation of  \eqref{13} is then
obtained by subdividing $[0,1]$ into $N$ subintervals of length $N^{-1}$
and letting $\omega$ be an $N$-step piecewise linear path $x=x_0\to
x_1\to\dots\to x_N=y$ for which
\begin{equation*}
\int_0^1|\dot{\omega}|^2\,dt = \frac{1}{ N}
\sum_{i=0}^{N-1}\left(\frac{x_{i+1}-x_i}{N^{-1}}\right)^2 =
N\sum_{i=0}^{N-1} (x_{i+1}-x_i)^2\,.
\end{equation*}
Setting 
\begin{equation*}
a^2 = \frac{T}{N}
\end{equation*}
we have 
\begin{align}\label{17}
H^a(x,y) &\equiv\;
\frac{a^2}{(2\pi a^2)^{d/2}} 
\sum_{N=1}^\infty\int\prod_{i=1}^{N-1}\frac{dx_i}{(2\pi
  a^2)^\frac{d}{2}} \exp\left(-\oh\sum_{i=0}^{N-1}
\frac{|x_{i+1}-x_i|^2}{a^2}- \oh m^2a^2N\right)\nonumber\\
& =\; a^2\sum_{N=1}^\infty (2\pi
  a^2N)^{-\frac{d}{2}} \exp\left(-\frac{|x-y|^2}{2a^2N}-\oh
m^2a^2N\right)\nonumber\\
& \to\; \int_0^\infty (2\pi
T)^{-\frac{d}{2}} \exp\left(-\frac{|x-y|^2}{2T}- \oh m^2T\right)\, 
dT\nonumber\\
& =\; G(x,y)
\end{align}
for $x\neq y$, as $a\to 0$. Hence, the function 
$H^a(x,y)$ defined here provides a
discrete approximation to $G(x,y)$. In the same way as for 
the hypercubic lattice
approximation we show in the Section 4 
that the measures on piecewise linear paths  
defined by the approximation
$H^a$ converge to a continuum path measure which attributes a proper
meaning to Eq.\ \eqref{13}.

\medskip

\section{The continuum measures and discrete approximations}

As noted in the previous section the appropriate space to integrate over in
Eqs.\ \eqref{3} and \eqref{13} consists of equivalence classes of paths under
reparametrizations. In this section we define those orbit spaces and the
relevant measures.

\subsection{Picewise linear paths}

It is convenient to start with Eq.\ \eqref{13} and for notational and technical
simplicity to consider first paths with only one fixed endpoint $x$. Let
$\Gamma(x)$ be the space consisting of pairs $(e,\omega)$ where
$e:[0,1]\to\mathbb R$ is a positive continuous function and
$\omega:[0,1]\to\mathbb R^d$ is continuous with $\omega(0)=x$. 
Let $\Diff_+[0,1]$ denote the set of all increasing diffeomorphisms of 
the unit interval.
As remarked in
the previous section there is a unique $\varphi\in\Diff_+[0,1]$ such that the
reparametrised pair $(e',\omega')$ defined by Eq.\ \rf{16} 
has $e'=T$ where $T$ is a constant. 
Hence we conclude that  
\begin{equation*}
\tilde\Gamma(x)\equiv\Gamma(x)/\Diff_+[0,1]=\mathbb R_+\times\Omega(x),
 \end{equation*}
 where $\Omega(x)$ denotes the set of continuous paths $\omega:[0,1]\to\mathbb
 R^d$  with $\omega(0)=x$.

Let us define a metric $\tilde d$ on $\tilde\Gamma(x)$ by
$$
\tilde
d((T,\omega),(T',\omega'))=|T-T'|+d(\omega,\omega'),
$$ 
where $d$ is the standard
uniform metric on  $\Omega(x)$ defined by
 \begin{equation*}
d(\omega,\omega')=\sup\{|\omega(s)-\omega'(s)|: s\in [0,1]\}.
 \end{equation*}
Equipped with $\tilde{d}$ the set  
$\tilde\Gamma(x)$ becomes a separable metric space. The discussion of
probability measures and their convergence properties is particularly
convenient on complete metric spaces (see, e.g., \cite{billingsley}). Since
$\Omega(x)$ with the metric $d$ is complete we can complete
$\tilde\Gamma(x)$ by adjoining $0\times\Omega(x)$. 
This will be assumed in the
folowing.  All measures on $\tilde\Gamma(x)$ that will be considered vanish
identically on $0\times\Omega(x)$.
 
On  $\Omega(x)$ we have the family of Wiener measures $W_x^t,\; t>0,$ defined
on the Borel subsets of  $\Omega(x)$. Here $t$ denotes the variance 
of the measure.  We note that
$W_x^t$ is uniquely defined by the characteristic functions of its finite
dimensional distributions which are 
given for $0<t_1<t_2<\dots<t_n\leq 1$ by
 \begin{align}\label{char1}
p^t_{t_1,\dots,t_n}(\xi_1,\dots,\xi_n) & =
\int \exp\left( i\xi_1\cdot\omega(t_1)+\dots+i\xi_n\cdot
\omega(t_n)\right)\, dW_x^t(\omega)\nonumber\\
& = \int\prod_{i=1}^n dx_i(2\pi t(t_i-t_{i-1}))^{-d/2}
\exp\left( -\frac{|x_i-x_{i-1}|^2}{2t(t_i-t_{i-1})}+i\xi_i \cdot x_i
\right) \nonumber\\ 
& =
\exp\left( -\frac{t}{2}\sum_{i=1}^n(t_i-t_{i-1})(\xi_i+\dots+\xi_n)^2
+ix\cdot (\xi_1+\dots+\xi_n)\right) \,,
 \end{align}
where $\xi_1,\dots,\xi_n\in\mathbb R^d$, $t_0=0$ and $x_0=x$.

For a  Borel set $B\subseteq\tilde\Gamma(x)$ we  let 
 \begin{equation*}
B_t=\{\omega : (t,\omega)\in B\}\quad\text{for}\quad t>0\;,
 \end{equation*}
and define the measure $W_x$ on $\tilde\Gamma(x)$ by
 \begin{equation*}
W_x(B)= \int_0^\infty e^{-\frac{1}{2} m^2t}\, W_x^t(B_t)dt\,.
 \end{equation*} 
The above definition requires $t\to W_x^t(B_t)$ to be a measurable function.
Rather than proving this directly we show that this must be the case by
giving an alternative definition of $W_x$. First,
let $x=0$ and consider the product 
$M$ of Lebesgue measure on $\mathbb R_+$ and $ W_0^1$ on $\Omega(0)$, i.e.,
 \begin{equation*}
dM(t,\omega)=  dtdW_0^1(\omega)\,.
 \end{equation*}
 Defining a homeomorphism $h$ of $\mathbb
R_+\times\Omega(0)$ onto itself by $h(t,\omega)= (t,t^{-\frac{1}{2}}\omega)$
and observing that 
 \begin{equation*}
W_0^t(A)= W_0^1(t^{-\frac{1}{2}}A)
 \end{equation*}
for Borel sets $A\subseteq\Omega(0)$, it follows that we have a measure
$W_0$ on $\tilde\Gamma(0)$ given by 
 \begin{equation*}
W_0(B)= \int_B e^{-\frac{1}{2} m^2t}d(M\circ h)(t,\omega)\,,
 \end{equation*}
where the measure $M\circ h$ on $\mathbb R_+\times\Omega(0)$ is
defined by $(M\circ h)(B)=M(h(B))$ for Borel sets  $B\subseteq\mathbb
R_+\times\Omega(0)$. This shows that $W_0$ is well defined. For arbitrary $x$
we obtain $W_x$ as the translation of $W_0$ by $x$.

To set up the discrete approximation to $W_x$, given by Eq.\ \eqref{17} for the
propagator, let $\tilde\Gamma_{a,N}(x)\subseteq\tilde\Gamma(x)$ 
be the set of pairs
$(T,\omega)$, where $T=a^2N$ and $\omega$ is an $N$-step piecewise linear
path $x=x_0\to x_1\to\dots\to x_N$ such that the step $x_{i-1}\to x_i$ is
parametrized linearly by the interval 
$[(i-1)/N,i/N]$. Define the
measure $W_{x,a,N}$ on $\tilde\Gamma(x)$ supported on 
$\tilde\Gamma_{a,N}(x)$ by  
 \begin{equation}\label{fin1}
dW_{x,a,N}(T,\omega)= \prod_{i=1}^N
dx_i(2\pi a^2)^{-\frac{d}{2}}\exp\left( -\frac{1}{2a^2}|x_i-x_{i-1}|^2\right)
 \,.
 \end{equation}
For $N=0$ we let $W_{x,a,0}$ be the Dirac measure at the trivial (constant)
path. The approximating measure $W_{x,a}$ on $\tilde\Gamma(x)$ is
supported on the set
$$
\tilde\Gamma_{a}(x) \equiv \bigcup_{N=0}^\infty \tilde\Gamma_{a,N}(x)
$$
and defined by 
 \begin{equation}\label{app1}
W_{x,a} = (1-e^{-\oh m^2a^2})\sum_{N=0}^\infty 
e^{-\frac{1}{2} m^2a^2N}W_{x,a,N}\;.
 \end{equation}
The normalization factor in Eq.\ \eqref{app1} has been chosen such that
$W_{x,a}$ is a probability measure, 
whereas the volume of $W_x$ is $W_x(\tilde\Gamma(x))=\frac{2}{m^2}$.
We prove the following result in the next section. 
\begin{theorem}\label{conv1}\rm 
\qquad$ W_{x,a} \to \frac{m^2}{2}W_x$ \quad as \quad $a\to 0\;.$
 \end{theorem}
Here and in the following convergence of measures is in the sense of {\it weak
convergence}, i.e., 
\begin{equation*}
\int f dW_{x,a} \to \int f dW_{x}\quad\text{as}\;\; a\to 0\;,
 \end{equation*}
for all bounded continuous functions $f$ on $\tilde\Gamma(x)$. 

\subsection{Lattice paths}

Next let us discuss the measure pertaining to Eq.\ \rf{3}. 
The relevant orbit space is now 
\begin{equation*}
\tilde\Omega(x)=\Omega(x)/\Diff_+[0,1]=\{[\omega]: \omega\in\Omega(x)\}\,,
 \end{equation*}
where $[\omega]=\{\omega\circ\varphi\mid
\varphi\in\Diff_+[0,1]\}$.  The quotient space 
$\tilde\Omega(x)$ inherits in a standard fashion a
pseudo-metric $\bar d$ from the metric $d$ on $\Omega(x)$, given by 
 \begin{equation*}
\bar d([\omega],[\omega'])=\inf\{d(\omega,\omega'\circ\varphi): \varphi\in
\Diff_+[0,1]\}\;. 
 \end{equation*}
Here the term pseudo-metric means that $\bar d([\omega],[\omega'])=0$ may occur
even if $[\omega]\neq[\omega']$.  For example, we have 
$\bar d([\omega],[\omega\circ f])=0$ whenever
$f:[0,1]\to[0,1]$ is a uniform limit of increasing diffeomorphisms. This
defect is eliminated by taking a further quotient setting
 \begin{equation*}
\bar\Omega(x)=\{\bar\omega : \omega\in\Omega(x)\}\;,
 \end{equation*}
where $\bar\omega =\{[\omega']: \bar d([\omega],[\omega'])=0\}$. Then $\bar d$
defines a metric on $\bar\Omega(x)$, and it is straightforward to verify
that $\bar\Omega(x)$ is a complete separable metric space. 

It is not hard to see that the same space
$\bar\Omega(x)$ results from the above construction if, e.g., we replace
$\Diff_+[0,1]$ by the group $\Homeo_+[0,1]$ of increasing homeomorphisms of
the unit interval.
Let us also note that evidently the quotient map
$\pi:\Omega(x)\to\bar\Omega(x)$ is continuous.

The measure $W_x$ on $\tilde\Gamma(x)=\mathbb R_+\times\Omega(x)$ constructed
in the previous subsection 
gives rise to a measure $V_x'$ on $\Omega(x)$ by integration over
the $t$-variable,
 \begin{equation*}
V_x'(A)= W_x(\mathbb R_+\times A)=\int_0^\infty e^{-\frac{1}{2}m^2t}W_x^t(A)dt
 \end{equation*}
for Borel sets $A\subseteq\Omega(x)$. Transporting this measure to
$\bar\Omega(x)$  by $\pi$ we obtain a measure $V_x$ given by 
 \begin{equation*}
V_x(\bar A) = V_x'(\pi^{-1}(\bar A))\;.
 \end{equation*}
This measure is defined on those sets
$\bar A$ for which $\pi^{-1}(\bar A)$ is a Borel set.  This $\sigma$-algebra
contains the Borel algebra of $\bar\Omega (x)$ since $\pi$ is 
continuous and we claim that the measure so defined is the appropriate one
to associate to Eq.\ \rf{3}.

In order to define the corresponding 
lattice approximation let $\Omega_{a,N}(x)$
denote the set of para\-metrized paths in $x+a\mathbb Z^d$ with $N$ steps, 
such
that the $i$th step is linearly parametrized by
$[\frac{i-1}{N},\frac{i}{N}]$.  Here $x$ is an arbitrary point in 
$\mathbb R^d$.  We 
let the discrete measure $V'_{x,a,N}$ on  $\Omega(x)$, 
supported on $\Omega_{a,N}(x)$ be defined by 
\begin{equation}\label{fin2}
V'_{x,a,N}(\omega) =
e^{-\beta_0N}\quad\text{for}\;\;\omega\in\Omega_{a,N}(x)\;, 
 \end{equation}
where $\beta_0 = \log 2d$, i.e., $V'_{x,a,N}$ is a normalized counting 
measure.

Furthermore, in correspondence with Eqs.\ \eqref{4} and \eqref{5} 
we define the 
measure $V'_{x,a}$ on $\Omega(x)$ supported on  
$$
\Omega_a(x) \equiv
\bigcup_{N=0}^\infty\Omega_{a,N}(x)
$$ 
by 
\begin{equation}\label{app2}
V'_{x,a} = (1-e^{-\ohd m^2a^2})\sum_{N=0}^\infty
e^{-\ohd m^2a^2N}V'_{x,a,N}\;. 
 \end{equation}
Here, $V'_{x,a,0}$ denotes the Dirac measure at the trivial path in
$\Omega(x)$, and the normalisation has been chosen such that $V'_{x,a}$ is a
probability measure. 

Similarly, we define 
$$
\bar\Omega_{a,N}(x) = \pi(\Omega_{a,N}(x))
$$
and 
$$
\bar\Omega_{a}(x) = \pi(\Omega_{a}(x)) =
 \bigcup_{N=0}^\infty\bar\Omega_{a,N}(x)\,.
$$
Correspondingly we define the transported
 measures  $V_{x,a,N}$ and $V_{x,a}$ given by
\begin{equation}\label{app3'}
V_{x,a,N}(\bar A) = V'_{x,a,N}(\pi^{-1}(\bar A))
 \end{equation}
and
\begin{equation}\label{app3}
V_{x,a}(\bar A)  = (1-e^{-\ohd m^2a^2})\sum_{N=0}^\infty e^{-\ohd 
m^2a^2N}V_{x,a,N}(\bar A)
 \end{equation}
for Borel sets $\bar A\subseteq \bar\Omega(x)$.
With these definitions we then have 
\begin{theorem}\label{conv2}\rm
\qquad$V_{x,a} \to \frac{m^2}{2}V_x$ \quad as \quad $a\to 0\;.$
 \end{theorem}
This result is proven in the subsequent section as a consequence of the
stronger result $V'_{x,a} \to \frac{m^2}{2}V'_x$ as $a\to 0\;.$ 

\subsection{Paths with two fixed endpoints}

Let us  briefly discuss paths with both endpoints $x,y$ fixed. It is
straightforward to introduce analogues to the spaces defined above for paths
with one fixed endpoint. We shall use the same notation except that $x$ is
everywhere replaced by $x,y$. On $\Omega(x,y)$ the family of Wiener measures
$W_{x,y}^t,\; t>0$, is defined by the characteristic functions  
 \bea
 q^t_{t_1,\dots,t_n}(\xi_1,\dots,\xi_n)  & = &
\int
\exp\left( i\xi_1\cdot\omega(t_1)+\dots+i\xi_n\cdot\omega(t_n)
\right)\,dW_{x,y}^t(\omega)\nonumber\\ 
& = & \int\prod_{i=1}^n dx_i(2\pi t(t_i-t_{i-1}))^{-d/2}
\exp\left(-\frac{|x_i-x_{i-1}|^2}{2t(t_i-t_{i-1})}+i\xi_i\cdot x_i\right)
\nonumber\\
&\times & (2\pi
t(1-t_{n}))^{-d/2}
\exp\left( -\frac{|y-x_{n}|^2}{2t(1-t_{n})}\right)\label{char2}\\ 
\lefteqn{=Z_{x,y}^t
\exp\left( -\frac{t}{2}\sum_{i=1}^n
(t_i - t_{i-1})(\xi_i+\dots+\xi_n)^2-
\left(\sum_{i=1}^nt_i\xi_i\right)^2
+i\sum_{i=1}^n(t_iy+(1-t_i)x)\xi_i\right)}~~~~~~~~~~~
~~~~~~~~~~~~&
&\nonumber
\eea 
where 
\begin{equation*}
Z_{x,y}^t=(2\pi t)^{-d/2}e^{-\frac{|x-y|^2}{2t}}\,,
\end{equation*}
the volume of $W_{x,y}^t$, is simply the heat kernel.

We then define the measure $W_{x,y}$ on $\tilde\Gamma(x,y)$ for $x\neq y$ by
 \begin{equation*}
W_{x,y}(B) = \int_0^\infty e^{-\frac{1}{2} m^2t}W_{x,y}^t(B_t)dt\;.
\end{equation*}
where $B\subseteq\tilde\Gamma(x,y)$ is a Borel set and $B_t
\subseteq\Omega(x,y)$ is defined as previously. The fact 
that this expression is
well defined is shown in a similar way as for $W_x$ by first noting that
 \begin{equation*}
W_{0,0}^t(A) = t^{-\frac{d}{2}}W_{0,0}^1(t^{-\frac{1}{2}}A)
\end{equation*} 
for Borel sets $A\subseteq\Omega(0,0)$ , and then using
 \begin{equation*}
W_{x,y}^t(A) = \exp\left(-\frac{|x-y|^2}{2t}\right) 
W_{0,0}^1(A-\omega_{x,y})\;,
\end{equation*} 
where $\omega_{x,y}$ is the linear path from $x$ to $y$ and $A$ is a Borel
subset of $\Omega(x,y)$. The last relation
is a direct consequence of Eq.\ \eqref{char2}.

Having defined $W_{x,y}$ the measures $V'_{x,y}$ and  $V_{x,y}$ are
defined in a similar way as  $V'_{x}$ and $V_{x}$.

The piecewise linear approximation is defined in analogy with
Eq.\ \eqref{app1} by 
\begin{equation}\label{app4}
W_{x,y,a} = (1-e^{-\oh m^2a^2})\sum_{N=0}^\infty 
e^{-\frac{1}{2} m^2a^2N}W_{x,y,a,N}\;,
 \end{equation}
where
 \begin{equation}\label{fin3}
dW_{x,y,a,N}(T,\omega)= \prod_{i=1}^{N}
dx_i(2\pi a^2)^{-\frac{d}{2}}\exp\left(
-\frac{1}{2a^2}|x_i-x_{i-1}|^2\right) 
 \end{equation}
for an $N$-step piecewise linear path $\omega:x=x_0\to x_1\to\dots\to
x_{N-1}\to x_N=y$.
Here $W_{x,y,a,1}$ is the Dirac measure
$\delta_{(1,\omega_0)}$, where $\omega_0$ is the linear path from $x$ to $y$, 
and $T=a^2N$ as before.  

Similarly, the hypercubic approximation is defined for $x\neq y$,
$x-y\in a\mathbb Z^d$, by 
\begin{equation}\label{app5}
V'_{x,y,a} = (1-e^{-m^2a^2})\sum_{N=1}^\infty e^{- m^2a^2N}V'_{x,y,a,N}\;,
 \end{equation}  
where
 \begin{equation}\label{fin4}
V'_{x,y,a,N}(\omega) =
a^{-d}e^{-\beta_0N}\quad\text{for}\;\;\omega\in\Omega_{a,N}(x,y)\;, 
 \end{equation}
and $V_{x,y,a}$ is obtained by transporting to $\bar\Omega(x,y)$ by the
quotient map $\pi$. 
%For  $W'_{x,y,a}$ and $\bar W_{x,y,a}$ it will be assumed
%in the following that the 
%lattice spacing $a$ is restricted to values for which $x-y\in a\mathbb Z^d$,
%since otherwise $\Omega_a(x,y)$ is empty. 
Since in all cases we are
interested in the limit $a\to 0$ we shall assume $0<a<1$.  

 It should be noted that in contrast to the case of paths with one fixed 
endpoint, the approximating 
measures defined here are not probability measures.
 The volume of $W_{x,y,a,N}$ is obtained by explicit computation and equals
\begin{equation}\label{norm}
Z_{x,y}^{a^2N} = (2\pi
a^2N)^{-\frac{d}{2}}e^{-\frac{|x-y|^2}{2a^2N}}\;, 
\end{equation}
which by insertion into \eqref{app4} immediately shows that the volume of $
W_{x,y,a}$ equals $\left(1-e^{-\oh m^2a^2}\right)a^{-2}H^a(x,y)$ 
and converges to
$\frac{m^2}{2}G(x,y)$ as $a\to 0$ according to Eq.\ 
\eqref{17}. Similarly, the
volume of $V_{x,y}$ equals $\left(1-e^{-\ohd m^2a^2}\right)
a^{-2} G^a(x,y)$ and converges to  $\frac{m^2}{2}G(x,y)$ 
as $a\to 0$ according to Eq.\ \eqref{12}.
On the other hand, the volume of $ W_{x,y}$ and of  $V_{x,y}$ both equal
$G(x,y)$.
 The convergence of volumes extends to the following result.

\begin{theorem}\label{conv3}\rm
\qquad $W_{x,y,a} \to \frac{m^2}{2}W_{x,y}$ \quad and \quad $V_{x,y,a}
\to \frac{m^2}{2} V_{x,y}$  \quad as \quad $a\to 0$ \quad for $x\neq y$\;.
 \end{theorem}
The proof is given in the next section.

\medskip
\section{Convergence of the approximations}

In a complete separable metric space $M$ there is a standard two-step procedure
for proving convergence of a family $m_a,\;a>0$, of Borel probability measures
to a measure $m$. The first step is to verify that $m_a,\;a>0$, is a {\it
  tight} (or {\it precompact}) family.   This means that 
for every $\eta>0$ there exists
a compact set  
$K\subseteq M$ such that $m_a(K)\geq 1-\eta$ for all $a>0$. The second
step is to show that 
\begin{equation}
\int_Mf\,dm_a\to\int_Mf\,dm\label{convergence}
\end{equation}
as $a\to 0$ for a collection of
functions that determine the measure in the sense that if the integrals of
these functions coincide for two measures then the measures coincide. Of
course, the first step is superfluous if one can establish the 
convergence \rf{convergence} for all
bounded continuous functions $f$. But this only happens rarely. Generally, the
first step ensures that every sequence $m_{a_n}$ from the given family has a
convergent subsequence, and the second step then implies that its limit is
independent of the chosen sequence or subsequence. For the spaces
$\Omega(x)$ and $\Omega(x,y)$ the second 
step can be
accomplished by proving convergence of the characteristic functions of the
finite dimensional distributions. For the spaces $\tilde\Gamma(x)$ and 
$\tilde\Gamma(x,y)$ 
a little more is required as we discuss below.

In the following four lemmas we show that the approximations introduced in
the previous section form tight families.

\begin{lemma}\label{tight1}\rm
 \qquad  $W_{x,a},\;0<a<1$, is a tight family of measures on
  $\tilde\Gamma(x)$.
\end{lemma}  

\begin{proof}  The following is an adaptation of the
corresponding argument for the piecewise linear approximations to the measure
$W_{x}^t$ (see \cite{billingsley}). 
According to the Arzela-Ascoli theorem the sets of compact
closure in $\Omega(x)$ are the equicontinuous ones. Defining the modulus of
continuity
\begin{equation*}
m(\omega,\delta)=\sup\{|\omega(s)-\omega(t)| :
|s-t|<\delta\}\quad\text{for}\quad \delta>0,\; \omega\in\Omega(x)\;,
\end{equation*}
it follows that complements to sets of the form
\begin{equation}\label{ccomp}
C=\bigcup_{n=1}^\infty \left\{ \omega : m(\omega,\delta_{n})>\frac{1}{n}
\right\}
\end{equation}
have compact closures in $\Omega(x)$ for an arbitrary sequence $\{ 
\delta_n\}$ of
positive numbers.   We observe that by Eq.\ \eqref{app1}
\begin{equation}\label{small}
W_{x,a}\left([t_0,+\infty )\times\Omega(x)\right)<\eta\quad\text{if}\quad
t_0>-m^{-2}\log\eta 
\end{equation}
for any $\eta>0$ and all $a>0$.   In order to prove the lemma it therefore
suffices to show that 
for any $\eta,\;\varepsilon,\;t_0 >0$ there exists a
$\delta>0$ such that  
\begin{equation}\label{eqc1}
W_{x,a}([0,t_0]\times\{\omega\in\Omega(x) : 
m(\omega,\delta)>\varepsilon\})<\eta 
\end{equation}
for all $a>0$. 
%Indeed, choosing $\delta_n$ in \eqref{ccomp} corresponding to 
%$\varepsilon=\frac{1}{n}$ and $2^{-n}\eta$
%instead of $\eta$ in \eqref{eqc1}, the measure of the complement to $C$ is
%larger than $1-\eta$.

 By Eq.\ \eqref{app1} it follows that Eq.\ \eqref{eqc1} holds if 
\begin{equation}\label{eqc2}
W_{x,a,N}\left(\{(a^2N,\omega)\in\tilde\Gamma(x) :
m(\omega,\delta)>\varepsilon\}\right)<\eta\quad\text{for}\quad a^2N\leq t_0\;. 
\end{equation} 
But for $a,N$ as in Eq.\ \eqref{eqc2} we have
\bea
 W_{x,a,N}(\{(a^2N,\omega) :  m(\omega,\delta)>\varepsilon\})
& = & W_{x,1,N}(\{(N,\omega) :
m(\omega,\delta)>\frac{\varepsilon}{a}\})\nonumber \\  
& \leq & W_{x,1,N}(\{(N,\omega) :
m(\omega,\delta)>\frac{\varepsilon\sqrt{N}}{\sqrt{t_0}}\})\,.  
\eea
Hence, it suffices to show, for given $ \eta,\varepsilon >0$, that 
\begin{equation}\label{eqc3}
W_{x,1,N}(\{(N,\omega) :
m(\omega,\delta)>\varepsilon\sqrt{N}\})<\eta\;,
\end{equation}
if $\delta$ is small enough.  This is a well known result (see,
e.g., \cite{billingsley} pp. 62-63). For later refrence 
we briefly recall the argument.

 First, note that since paths contributing to \eqref{eqc3} are linear on
 each interval $[\frac{i-1}{N},\frac{i}{N}]$ we have
\begin{equation*}
m_{N}(\omega,\delta)\equiv\max\left\{|\omega(\frac{i}{N})-
\omega(\frac{j}{N})| :
 0\leq i,j\leq N,\;|\frac{i}{N}-\frac{j}{N}|<\delta \right\}\geq
 \frac{1}{3}m(\omega,\delta)
\end{equation*} 
for $N\geq\delta^{-1}$. Note also that by uniform continuity of
$\omega\in\Omega(x)$ the inequality \eqref{eqc3} is fulfilled for sufficiently
small $\delta$  for each individual $N$, so we need not worry about small
$N$. Hence we may replace $ m(\omega,\delta)$ in
\eqref{eqc3} by $m_{N}(\omega,\delta)$, and we may assume
$\delta=M^{-1}$, where $M\in \mathbb N$ and $N\geq M$. 

Next, given $N\geq M$, we choose integers $0=k_0<k_1<\dots<k_M=N$
such that any subinterval $[\frac{i}{N},\frac{j}{N}]$ 
of $[0,1]$ of length $\leq\delta$ is contained in one of
the intervals $[\frac{k_l}{N},\frac{k_{l+2}}{N}]$ and such that the latter
intervals are all of length  $\leq 3\delta$. It follows that 
\begin{align*}
& W_{x,1,N}\left(\{(N,\omega) : m(\omega,\delta)>\varepsilon\sqrt{N}\}\right)\\
\leq &\; \sum_{l=0}^{M-2} W_{x,1,N}\left(\left\{
(N,\omega) : \underset{k_l\leq k\leq
  k_{l+2}}\max
|\omega(\frac{k_l}{N})-\omega(\frac{k}{N})|>\frac{\varepsilon}{6}\sqrt{N}
\right\}\right)\\
\leq &\; \sum_{l=0}^{M-2} W_{x,1,N}\left(\left\{(N,\omega) : 
\underset{k_l\leq k\leq
  k_{l+2}}\max
|\omega(\frac{k_l}{N})-\omega(\frac{k}{N})|>\frac{
\delta^{-\frac{1}{2}}\varepsilon}{6\sqrt{3}}
\sqrt{k_{l+2}-k_l}\right\}\right)\;.
 \end{align*}
Due to statistical independence of the steps in $\omega$ and translation
invariance, we have
\begin{align*}
& W_{x,1,N}\left(\left\{(N,\omega)\in\tilde\Gamma(x) : 
\underset{k_l\leq k\leq k_{l+2}}
\max|\omega(\frac{k_l}{N})-\omega(\frac{k}{N})|>\alpha\right\}\right)\\
= &\; W_{0,1,k_{l+2}-k_l}\left(\left\{(k_{l+2}-k_l,\omega)\in\tilde
\Gamma(0) : \underset{k_l\leq k\leq
  k_{l+2}}\max|\omega(\frac{k}{k_{l+2}-k_l})|>\alpha\right\}\right) 
\end{align*}
for $\alpha >0$. Combining this with the previous inequality we
conclude that it is sufficient to show for given 
$\eta>0$ the existence of $\delta>0$ such that for all $N\in\mathbb N$
\begin{equation*}
\delta^{-1}W_{0,1,N}\left(\left\{(N,\omega)
: \max\{|\omega(\frac{i}{N})|\mid 0\leq i\leq N\}
>\delta^{-\frac{1}{2}}\sqrt{N}\right\}\right)<\eta\;. 
\end{equation*}
This inequality is a consequence of the Chebychev
inequality and the uniform boundedness in $N$ of the moments of
$N^{-\frac{1}{2}}|\omega(1)|$ with respect to the measure 
$W_{0,1,N}$. The details may be found in
\cite{billingsley}.
\end{proof}

\medskip

\begin{lemma}\label{tight2}\rm
$V'_{x,a},\;0<a<1$, is a tight family of measures on $\Omega(x)$.
\end{lemma}

\begin{proof} This is obtained by an argument similar to the one
given above. First, applying the Arzela-Ascoli theorem one concludes that it
is sufficient to prove for given $\eta,\varepsilon>0$ that
\begin{equation}\label{eqc4}
V'_{x,a}(\{\omega\in\Omega(x) :
m(\omega,\delta)>\varepsilon\})<\eta\;.
\end{equation}
for small enough $\delta$.
Second, since the contribution of terms with $a^2N\geq t_0$ in \eqref{app2} is
less than or equal to $\frac{\eta}{2}$ if $t_0\geq -m^2\log\frac{\eta}{2}$ we
conclude as above that it is sufficient to show the existence of a 
$\delta>0$ such
that for all $N\in\mathbb N$
\begin{equation}\label{eqc5}
V'_{x,1,N}(\{\omega\in\Omega(x) :
m(\omega,\delta)>\varepsilon\sqrt N\})<\eta\;.
\end{equation}
The proof of this fact parallels the one for piecewise linear paths referred to
above, and uses only the statistical independence of the steps in a path
together with the uniform boundedness in $N$ of the moments of
$N^{-\frac{1}{2}}|\omega(1)|$ with respect to the measure 
$V'_{x,1,N}$. We
omit the details of the argument. 
\end{proof}

\begin{lemma}\label{tight3}\rm
$W_{x,y,a},\;0<a<1$, is a tight family of measures on $\tilde\Gamma(x,y)$ for
$x\neq y$ .
\end{lemma}

\begin{proof}:  By definition tightness of the family $W_{x,y,a},\;a>0$, means
tightness of the corresponding famify of normalized measures. Since, however,
the volume of  $W_{x,y,a}$ converges to the volume of 
$\oh m^2W_{x,y}$ as $a\to 0$,
as noted previously, we need not worry about normalisation.

We note first that the volume of $W_{x,y,a,N}$ given by \eqref{norm} is
uniformly bounded in $a$ and $N$. Hence, in \eqref{app4} the sum over $N\leq
s_0a^{-2}$ or over $N\geq t_0a^{-2}$ can be made arbitrarily small for
sufficiently small $s_0$ or sufficiently large $t_0$, respectively. By the same
arguments as in the first part of the proof of Lemma \ref{tight1} it is
sufficient to demonstrate the existence of a $\delta>0$ such that  
\begin{equation}\label{eqc6}
W_{x,y,a,N}(\{(a^2N,\omega)\in\tilde\Gamma(x,y) :
m(\omega,\delta)>\varepsilon\})<\eta\quad\text{for}\quad s_0\leq a^2N\leq
t_0\;,  
\end{equation} 
for given $\eta,\epsilon,s_0,t_0>0$.  

We may as before replace $m(\omega,\delta)$ by
$m_N(\omega,\delta)$. Assuming $\delta<\frac{1}{3}$ and setting
$N_1=[\frac{2}{3}N]+1,\;N_2=[\frac{1}{3}N]$ 
(where $[\alpha]$ denotes the integer
part of $\alpha$), any subinterval 
of $[0,1]$ of lenghth $\delta$ is contained in
either $[0,N_1/N]$ or in  $[N_2/N,1]$. Hence, we have 
\begin{align}\label{18}
& W_{x,y,a,N}(\{(a^2N,\omega) :
m_N(\omega,\delta)>\varepsilon\})\\
\leq &\; W_{x,y,a,N}(\{(a^2N,\omega) :
m^1_N(\omega,\delta)>\varepsilon\})+W_{x,y,a,N}(\{(a^2N,\omega) :
m^2_N(\omega,\delta)>\varepsilon\})\;,\nonumber 
\end{align}
 where we have set
\begin{equation*}
m^1_{N}(\omega,\delta)=\max\{|\omega(\frac{i}{N})-\omega(\frac{j}{N})| :
 0\leq i,j\leq N_1,\;|\frac{i}{N}-\frac{j}{N}|<\delta\} 
\end{equation*}
and
\begin{equation*}
m^2_{N}(\omega,\delta)=\max\{|\omega(\frac{i}{N})-\omega(\frac{j}{N})| :
 N_2\leq i,j\leq N,\;|\frac{i}{N}-\frac{j}{N}|<\delta\} \;.
\end{equation*}
By definition of $W_{x,y,a,N}$ we have
\begin{align}\label{est}
&W_{x,y,a,N}(\{(a^2N,\omega): m^1_{N}(\omega,\delta)>\varepsilon\})\nonumber\\
= &\;\int_{\mathbb R^d} du\, 
Z_{u,y}^{a^2(N-N_1)}W_{x,u,a,N_1}(\{(a^2N_1,\omega):
 m_{N_1}(\omega,\delta)>\varepsilon\}) \;.
\end{align}
Here $a^2(N-N_1)\geq a^2(\frac{1}{3}N-1)\geq \frac{1}{6}s_0$ (assuming $N\geq
6$) so
$$
Z_{u,y}^{a^2(N-N_1)}\leq (\frac{1}{3}\pi s_0)^{-\frac{d}{2}}.
$$
 Using this
estimate together with 
\begin{equation*}
dW_{x,a,N}(a^2N,\omega)= d\omega(1)
dW_{x,\omega(1),a,N}(\omega)\quad\text{for}\quad \omega\in\Omega(x)
\end{equation*}
in Eq.\ \eqref{est} we obtain
\begin{align*}
& W_{x,y,a,N}(\{(a^2N,\omega): m^1_{N}(\omega,\delta)>\varepsilon\})\\
\leq\;&  (\frac{1}{3}\pi s_0)^{-\frac{d}{2}}W_{x,a,N_1}(\{(a^2N_1,\omega) :
m_{N_1}(\omega,\delta)>\varepsilon\})\;. 
\end{align*}
Finally, using $a^2N_1\leq \frac{2}{3}t_0$ we conclude from the proof of
Lemma \ref{tight1} that\linebreak $W_{x,a,N_1}(\{(a^2N_1,\omega) :
m_{N_1}(\omega,\delta)>\varepsilon\})$ 
can be made arbitrarily small for $a^2N\leq
t_0$ if $\delta$ is chosen small enough.

The second term in \eqref{18} can be treated similarly,  and the lemma is
proven. 
%Next we make use of the identity
%\begin{equation*}
%W_{x,y,a,N}(a^2N\times B) = e^{-\frac{|x-y|^2}{2a^2N}}W_{0,0,a,N}(a^2N\times (B-\omega_{x,y}))\;, 
%\end{equation*}
%where $\omega_{x,y}$ is the linear path from $x$ to $y$. Since
%$m(\omega,\delta)\leq m(\omega-\omega_{x,y},\delta)+\delta|x-y|$ we obtain for
%$a,N$ as in \eqref{eqc6} and $\delta<\frac{\epsilon}{2|x-y|}$ 
%\begin{align*}
%& W_{x,y,a,N}(a^2N\times\{\omega\in\Omega(x,y)\mid
%m(\omega,\delta)>\varepsilon\})\\
%\leq &  e^{-\frac{|x-y|^2}{2a^2N}}W_{0,0,a,N}(a^2N\times\{\omega\in\Omega(0,0)\mid
%m(\omega,\delta)>\frac{\varepsilon}{2}\})\\
%= & a^{-d}e^{-\frac{|x-y|^2}{2a^2N}}W_{0,0,1,N}(N\times\{\omega\in\Omega(0,0)\mid
%m(\omega,\delta)>\frac{\varepsilon}{2a}\})\\
%\leq & a^{-d}e^{-\frac{|x-y|^2}{2a^2N}}W_{0,0,1,N}(N\times\{\omega\in\Omega(0,0)\mid
%m(\omega,\delta)>\frac{\varepsilon}{2t_0}\sqrt{N}\})\;.
%\end{align*}
%
%In order to estimate the last expression we first replace $m(\omega,\delta)$
%by $m_N(\omega,\delta)$ and pick a subdivision
%$[\frac{k_0}{N},\frac{k_{1}}{N},\dots,\frac{k_M}{N}]$ as in the proof of Lemma
%\ref{tight1}. We then have 
%\begin{align*}
%& W_{0,0,1,N}(N\times\{\omega\in\Omega(0,0)\mid
%m_N(\omega,\delta)>\varepsilon\})\\
%\leq & \sum_{l=0}^{M-2} W_{0,0,1,N}(N\times A_\varepsilon(k_l,k_{l+2}))
%\end{align*}
%and 
%\begin{align*}
% W_{0,0,1,N}(N\times A_\varepsilon(k_l,k_{l+2})) &=\int W_{x,z,1,k_{l+2}}(k_{l+2}\times
% A_\varepsilon(k_l,k_{l+2}))N_{z,y}^{N-k_{l+2}}dz \\
%&\leq (2\pi(N-k_{l+2}))^{-\frac{d}{2}}W_{x,1,k_{l+2}}(k_{l+2}\times
% A_\varepsilon(k_l,k_{l+2}))\;.
%\end{align*}
\end{proof}

\begin{lemma}\label{tight4}\rm
$V'_{x,y,a},\;0<a<1$, is a tight family of measures on $\Omega(x,y)$ for
$x\neq y$ .
\end{lemma}

\begin{proof} Only a few modifications of the previous proof are needed. 

For the volume $Z_{x,y,1,N}$ of $W_{x,y,1,N}$ we have the following result,
which is rather easily derived from its Fourier representation (see, e.g.,
\cite{spitzer} pp. 76-77]):
\begin{equation*}
\lim_{N\to\infty}\biggl((2\pi N/d)^{\frac{d}{2}}Z_{x,y,1,N}-e^{-\frac{|x-y|^2}{2N/d}}\biggr)=0
\end{equation*}
uniformly in $x-y\in\mathbb Z^d$. For the volume $Z_{x,y,a,N}$ of $W_{x,y,a,N}$
this means 
\begin{equation}\label{lim}
\lim_{N\to\infty}\biggl((2\pi a^2 N/d)^{\frac{d}{2}}Z_{x,y,a,N}-e^{-\frac{|x-y|^2}{2Na^2/d}}\biggr)=0
\end{equation}
uniformly in  $x-y\in a\mathbb Z^d$ and $0<a<1$.

As a first consequence of \eqref{lim} we note that $Z_{x,y,a,N}$ is uniformly
bounded in $a,N$ for $a^2N\geq t_0$ for any given $t_0>0$ large enough. It
follows that in Eq.\ \eqref{app4} the sum over $N\geq t_0a^{-2}$ can be made
arbitrarily small (when applied to any Borel set in $\Omega(x,y)$) by choosing
$t_0$ large enough. 

A second consequence is that for any $s_0>0$
\begin{align*}
&(1-e^{-\ohd m^2a^2})\sum_{a^2N\geq s_0}^\infty e^{-\ohd m^2a^2N}
Z_{x,y,a,N}\\
\to\;&\frac{m^2}{2d}\int_{s_0}^\infty
Z^{t/d}_{x,y}\, e^{-\frac{1}{2d}m^2t}dt =\;\frac{m^2}{2}\int_{s_0/d}^\infty
Z^{t}_{x,y}e^{-\frac{1}{2}m^2t}dt
\end{align*}
as $a\to 0$. On the other hand, as we know from Section 2, 
\begin{equation*}
(1-e^{-\frac{1}{2d}m^2a^2})\sum_{N=0}^\infty e^{-\frac{1}{2d}m^2a^2N}
Z_{x,y,a,N}\;\to\;\frac{m^2}{2}G(x,y)=\frac{m^2}{2}\int_{0}^\infty
Z^t_{x,y}e^{-\frac{1}{2}m^2t}dt 
\end{equation*}
as $a\to 0$. Hence we conclude that the sum in Eq.\ 
\eqref{app4} over $a^2N\leq
s_0$ can be made arbitrarily small 
for all $a<a_0$ for some $a_0>0$. Replacing
$s_0$ by $\min\{s_0,a_0^2\}$ we can arrange that $a_0=1$. 

It now follows as in the previous proof that it suffices to show for given
$\eta,\varepsilon,s_0,t_0>0$ that there exists $\delta>0$ such that 
\begin{equation*}
V'_{x,y,a,N}(\{\omega :
m_N(\omega,\delta)>\varepsilon\})<\eta\quad\text{for}\quad s_0<a^2N<t_0\;.
\end{equation*} 

Following the proof of Lemma 4.3 we have the estimate
 \begin{align}\label{19}
& V'_{x,y,a,N}(\{(a^2N,\omega) :
m_N(\omega,\delta)>\varepsilon\})\nonumber\\
\leq &\; V'_{x,y,a,N}(\{(\omega) :
m^1_N(\omega,\delta)>\varepsilon\})+V'_{x,y,a,N}(\{(\omega) :
m^2_N(\omega,\delta)>\varepsilon\})\;. 
\end{align}
By definition of $ V'_{x,y,a,N}$ we can write
\begin{align*}
&V'_{x,y,a,N}(\{(\omega): m^1_{N}(\omega,\delta)>\varepsilon\}\\
= &\;\sum_{u\in a\mathbb Z^d}Z_{u,y,a,N-N_1}V'_{x,u,a,N_1}(\{(\omega):
 m_{N_1}(\omega,\delta)>\varepsilon\} \;.
\end{align*}

From Eq.\ \eqref{lim} and $a^2(N-N_1)\geq \frac{1}{6}s_0$ it follows that
$Z_{u,y,a,N-N_1}$ is uniformly bounded in $x,y,a,N$ for $a^2N\geq s_0$ and
$n\geq N_0$ for some $N_0\in\mathbb N$. Letting $C$ denote such an upper
bound, we have 
\begin{align}
& V'_{x,y,a,N}(\{(\omega) : m^1_{N}(\omega,\delta)>\varepsilon\}\nonumber\\
\leq &\; C \sum_{u\in a\mathbb Z^d} V'_{u,x,a,N_1}((\{(\omega)\:
 m_{N_1}(\omega,\delta)>\varepsilon\}\nonumber\\
= &\;C V'_{x,a,N_1}((\{(\omega): m_{N_1}(\omega,\delta)>\varepsilon\}\;.
\label{qqq}
\end{align}
From the proof of Lemma 4.2 it now follows as above that the right
hand side of \eqref{qqq} can be made arbitrarily small for $a^2N\leq t_0$
if $\delta$ is chosen 
sufficiently small. Estimating the second term in \eqref{19}
similarly and noting again that the case $N<N_0$ can be taken care of
separately, the proof of the lemma is complete. 
\end{proof}

We are now ready to give proofs of the convergence theorems stated in 
Section 3. In view of the preceding lemmas and the remarks at the beginning of
this section it is sufficient in each case to prove convergence on a measure
determining class of functions.

\noindent
{\it Proof of Theorem \ref{conv1}:}\; For $0<t_1<\dots<t_n\leq 1$ and
  $s\in\mathbb R,\;\xi_1,\dots,\xi_n\in\mathbb R^d$ we define the
  characteristic function $p_{a;t_1,\dots,t_n}$ of $W_{x,a}$ by
\begin{equation}\label{char3}
p_{a;t_1,\dots,t_n}(s,\xi_1,\dots,\xi_n)= \int_{\tilde
\Gamma(x)}e^{i(st+\xi_1\cdot \omega(t_1)+
\cdots+\xi_n\cdot \omega(t_n))}dW_{x,a}(t,\omega)\;,
\end{equation}
and similarly the characteristic function $p_{t_1,\dots,t_n}$ of $W_{x}$.
We claim it is sufficient to show that $p_{a;t_1,\dots,t_n}\to\frac{m^2}{2}
p_{t_1,\dots,t_n}$ pointwise as $a\to 0$, for arbitrary $0<t_1<\dots<t_n\leq
1$. In order to see this, it is enough to verify that the measure $W_x$ on
$\tilde\Gamma(x)$ is determined by its characteristic functions. Let $f$ be a smooth function on $\mathbb
R_+\mathbb \times 
R^{nd}$ with compact support. Multiplying  $p_{t_1,\dots,t_n}$ by
the Fourier transform of $f$ at $(s,\xi_1,\dots,\xi_n)$ and integrating over
$(s,\xi_1,\dots,\xi_n)\in\mathbb R^{nd+1}$ gives by Fubini's theorem
$\int_{\tilde\Gamma(x)}f(t,\omega(t_1),\dots,\omega(t_n))dW_{x,a}(t,\omega)$. A
simple limiting argument then shows that measures of sets of the form
$\{(t,\omega): (t,\omega(t_1),\dots,\omega(t_n))\in C\}$, where
$C\subseteq\mathbb R_+\times\mathbb R^{nd}$ is closed, are determined by the
characteristic functions. Since sets of this form generate the Borel algebra
in $\tilde\Gamma(x)$ the claim follows.
 
Given $N$ let $1\leq N_1\leq\dots\leq N_n\leq N$ be such that $t_i\in
]\frac{N_i-1}{N},\frac{N_i}{N}]$ and set $t'_i=\frac{N_i}{N}$. By an explicit
computation, replacing the intermediate times $t_i$ by $t_i'$ in the
piecewise linear paths, one finds 
\begin{equation*}
 \int_{\tilde\Gamma(x)}
e^{i(st+\xi_1\cdot \omega(t_1)+\cdots+\xi_n\cdot 
\omega(t_n))}dW_{x,a,N}(t,\omega)= 
C_{a,N}\, p^{a^2N}_{t'_1,\dots,t'_n}(s,\xi_1,\dots,\xi_n).
\end{equation*}  
The quantity $C_{a,N}$  which depends on the time differences
$t_i-t_i'$ tends to 1 uniformly in $N$ as $a\to 0$. 
Using the expression
\eqref{char1} for 
$p^t_{t_1,\dots,t_n}(s,\xi_1,\dots,\xi_n)$ it follows easily
that
\begin{align*}
p_{a;t_1,\dots,t_n}(s,\xi_1,\dots,\xi_n)\;=\;
&(1-e^{-\frac{1}{2}m^2a^2})\sum_{N=0}^\infty
C_{a,N}e^{-\frac{1}{2}m^2a^2N}e^{isa^2N}p^{a^2N}_{t'_1,
\dots,t'_n}(s,\xi_1,\dots,\xi_n)\\
\to\;&\frac{m^2}{2}p_{t_1,\dots,t_n}(s,\xi_1,\dots,\xi_n)
\end{align*} 
as $a\to 0$.
%\end{proof}
\hfill $\Box$

\medskip

\noindent
{\it Proof of Theorem \ref{conv2}:}\; We do this by proving the
  stronger result that $V'_{x,a}\to\frac{m^2}{2}V'_x$ as $a\to 0$. 
It 
  follows by the same argument as given in the beginning of the previous
  proof that it is enough to prove
  convergence of the characteristic function
\begin{align*}
p'_{a;t_1,\dots,t_n}(\xi_1,\dots,\xi_n)\;=\;&
\int_{\Omega(x)}e^{i(\xi_1\cdot\omega(t_1)+\cdots+\xi_n\cdot
\omega(t_n))}dV'_{x,a}(t,\omega)\\
=\;& (1-e^{-\frac{1}{2}m^2a^2})\sum_{N=0}^\infty
e^{-\frac{1}{2}m^2a^2N}\sum_{\omega\in\Omega_{a,N}(x)}
e^{-\beta_0N}e^{i(\xi_1\cdot\omega(t_1)+\dots+\xi_n\cdot\omega(t_n))}
\end{align*}
to $\frac{m^2}{2}p_{t_1,\dots,t_n}$ as $a\to 0$ for arbitrary
$0<t_1<\dots<t_n\leq 1$. Furthermore, we can assume $x=0$, since translation
by $x$ only gives rise to a factor $e^{ix\cdot (\xi_1+\dots+\xi_n)}$ in the
characteristic functions. Defining $N_i$ and $t'_i$ as in the preceding proof
we have 
\begin{align*}
&p'_{a;t_1,\dots,t_n}(\xi_1,\dots,\xi_n)\\
=\;&
 (1-e^{-\frac{1}{2d}m^2a^2})\sum_{N=0}^\infty
e^{-\frac{1}{2d}m^2a^2N}\prod_{i=1}^n\biggl(\frac{1}{d}\sum_{\nu=1}^d\cos
a(\xi_i+\dots+\xi_n)_{\nu}\biggr)^{N_i-N_{i-1}}\\
=\;&(1-e^{-\frac{1}{2d}m^2a^2})\sum_{N=0}^\infty
e^{-\frac{1}{2d}m^2a^2N}\prod_{i=1}^n\biggl(\frac{1}{d}\sum_{\nu=1}^d\cos
a(\xi_i+\dots+\xi_n)_{\nu}\biggr)^{(t'_i-t'_{i-1})N}\;,
\end{align*}
where $\nu$ labels the components of the $\xi$-variables and $N_0=0$. Finally,
using 
\begin{equation*}
\biggl(\frac{1}{d}\sum_{\nu=1}^d\cos
a(\xi_i+\dots+\xi_n)_{\nu}\biggr)^{\frac{s}{a^2}}\;\to\;e^{-\frac{s}{2d}(\xi_i+\dots+\xi_n)^2}
\end{equation*}
as $a\to 0$, an application of the dominated convergence theorem shows that
\begin{equation*}
p'_{a;t_1,\dots,t_n}(\xi_1,\dots,\xi_n)\;\to\;\frac{m^2}{2}\int_0^\infty dt\,
e^{
-\frac{1}{2}m^2t}\exp\left(-\frac{t}{2}\sum_{i=1}^n(t_i-t_{i-1})(\xi_i+\dots+
\xi_n)^2\right)
\end{equation*}
as $a\to 0$, which is the desired result.
\hfill $\Box$

\medskip

\noindent
{\it Proof of Theorem \ref{conv3}:}\; The convergence
  $W_{x,y,a}\to W_{x,y}$ follows by essentially the same proof as of Theorem
  \ref{conv1}.
Similarly the convergence  $V_{x,y,a}\to V_{x,y}$ is obtained by
trivial modifications of the proof of Theorem 
\ref{conv2}. Details are left to the reader.

\hfill $\Box$

\section{Cylinder sets}

In this section we define a class of sets of geometric paths which 
generate the Borel algebra and play a role similar to the one 
played by cylinder sets in the theory of parametrized paths.  We will
see in the next section that the measure of these sets can be calculated 
in a particularly simple way.  

A natural condition to put on a parametrized 
path $\omega$ is that the path be located in a particular subset $A$ of 
$\mathbb R^d$ at a given time $t$, i.e., 
$\omega (t)\in A$.  For geometric 
paths a condition of this type 
is meningless but a similar one which has a well
defined meaning is the condition that a geometric path $\bar\omega $ 
hit a set $A$.  This means that $\bar\omega \cap A\neq\emptyset$,
i.e., if $\omega$ is a parametrization of $
\bar\omega $ then there is a time $t$ such that
$\omega (t)\in A$.
More generally, we can require that a geometric path hit a number of sets
in a particular order and/or stay away from other sets.  Below we define
a certain class of sets defined by such conditions.  Other definitions 
are possible but we find this class simple to work with.

We consider paths with two fixed endpoints $x$ and $y$.  
Let $A_1,\ldots, A_n$ be subsets of $\mathbb R ^d$ and let $\bar\omega\in
\bar\Omega (x,y)$ with parametrization $\omega: [0,1]\to \mathbb R^d$.
Define
\bea
\tau_1 & = & \sup\{ t\geq 0 :\omega([0,t])\subseteq A_1\} \nonumber\\
\tau_2 & = & \sup\{ t\geq \tau_1 :\omega([\tau_1,t])\subseteq A_2\}
\nonumber\\
       & \vdots &\nonumber \\ 
\tau_n & = & \sup\{ t\geq \tau_{n-1} :\omega([\tau_{n-1},t])\subseteq A_n\},
\nonumber
\eea
where by convention $\sup\emptyset =1$.
We then define $Z(A_1,\ldots ,A_n)$ as the set of all geometric paths
$\bar\omega\in\bar\Omega (x,y) $ such that
$$
\tau_1<\tau_2<\ldots <\tau_{n-1}<\tau_n=1.
$$
This defining property is easily seen to be independent of the 
parametrization $\omega$ chosen for $\bar\omega$.  In fact, $\bar\omega
\in Z(A_1,\ldots ,A_n)$ exactly if it starts at $x\in A_1$, stays inside $A_1$ 
until it leaves $A_1$ at a point $x_1=\omega (\tau_1)\in A_2$, then stays in 
$A_2$ until it leaves at a point $x_2=\omega (\tau_2)\in A_3$ and so on
until it leaves $A_{n-1}$ at a point $x_{n-1}=\omega (\tau_{n-1})\in A_n$ and
then finally 
stays in $A_n$ until it ends at $y\in \bar A_n$.  The values of the 
escape times $\tau _i$ depend of course on the parametrization but
their ordering and the points $x_i$ are independent of parametrization.

\medskip

\noindent
{\bf Proposition 5.1.}  Let $A_1,\ldots, A_n \subseteq \mathbb R ^d$ be 
open sets such that $x\in A_1$ and $y\in A_n\setminus\bar A_{n-1}$.  
Furthermore, assume
\begin{equation}
A_{i-1}\cap \pa A_i\cap A_{i+1} =\emptyset \label{star} 
\end{equation}
for $i=2,\ldots ,n-1$.  Then $Z(A_1,\ldots ,A_n)$ is an open subset of
$\bar\Omega (x,y)$.

\medskip
\noindent
{\it Proof.}  Let $\bar\omega\in Z(A_1,\ldots ,A_n)$.  Choose a parametrization
$\omega$ for $\bar\omega$.  Since the sets $A_i$ are open we can choose
$s_i<\tau_i$ such that $\omega([s_i,\tau_i])\subseteq A_{i+1}$, see Fig.\ 1.  
By the 
definition of the $\tau_i$'s it follows that $\omega ([s_i,s_{i+1}])\subseteq
 A_{i+1}$ for $i=0,1,\ldots ,n-1$, setting $s_0=0$.

\begin{figure}[hbt]
  \begin{center}
    \psfrag{tau1}{$\tau_1$}
    \psfrag{tau2}{$\tau_2$}
    \psfrag{x}{$x$}
    \psfrag{y}{$y$}
\psfrag{A1}{$A_1$}
\psfrag{A2}{$A_2$}
\psfrag{A3}{$A_3$}
    \psfrag{s1}{$s_1$}
    \psfrag{s2}{$s_2$}
    \includegraphics[width=12cm]{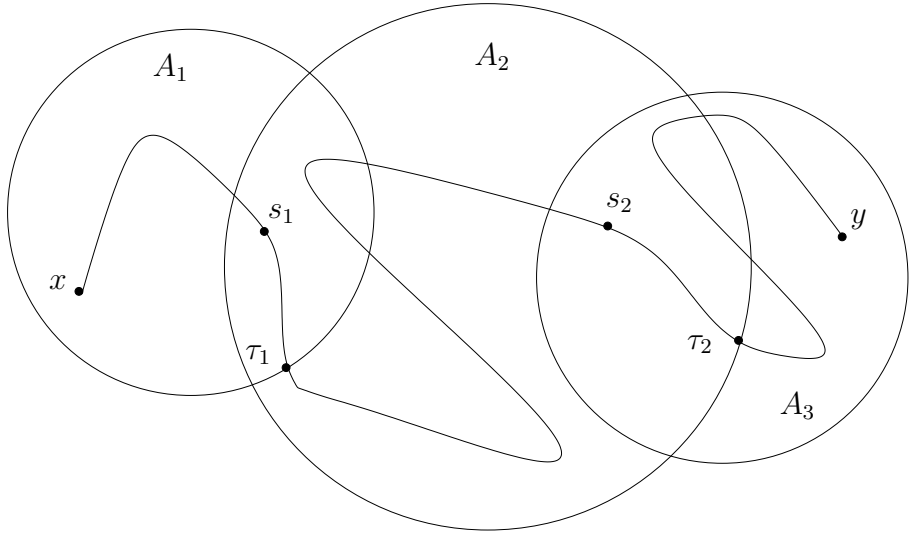}
    \caption{An illustration of the times $s_i$ and $\tau_i$ in 
the case of $n=3$.}
    \label{fig1}
  \end{center}
\end{figure}

Let $r_i>0$ be the distance from the compact set $\omega ([s_i,s_{i+1}])$ 
to the boundary of $A_{i+1}$, $i=0,1,\ldots ,n-1$, and set $r=\min_i r_i$.
Now take a geometric path $\bar\omega '$ at a distance smaller than $r$ from
$\bar\omega$.  Then there exists a parametrization $\omega ':[0,1] \to
\mathbb R^d$ of
$\bar\omega '$ such that
$$
\sup_{t\in [0,1]}|\omega (t)-\omega '(t)|<r.
$$
In particular it follows that 
$$
\omega '([s_i,s_{i+1}])\subseteq
 A_{i+1}.
$$

By the assumption \rf{star} we may from the outset choose the $s_i$'s such that
$\omega (s_{i+1})\notin A_i$.  Choosing $r$ smaller, if necessary, we can also 
assume that $r$ is smaller than the smallest of the distances from
$\omega (s_{i+1})$ to $A_i$, $i=1,2,\ldots ,n$.  Hence, 
$\omega '(s_{i+1})\notin A_i$.  
On the other hand $\omega '(s_i)\in A_i$ so $\omega '$ 
leaves the set $A_i$ at a time $\tau_i'\in [s_i,s_{i+1}]$.  It follows that 
$\tau_1'<\tau_2'<\ldots \tau_n'=1$ so $\bar\omega '\in  Z(A_1,\ldots ,A_n)$.
\hfill $\Box$

\medskip

The condition \rf{star} was essential in the above argument because otherwise
the paths might never enter the interior of $A_i\setminus A_{i-1}$ for 
some $i$.  But \rf{star}  can 
be replaced by a weaker condition as we now explain.  

Let $A$ be an open set in $\mathbb R^d$.
We say that a geometric path $\bar\omega$ is {\it tangent to 
the boundary of $A$} at $x\in \pa A$ if there is a parametrization $\omega$ 
of $\bar\omega$ such that $\omega (t_0)=x$ and there is an 
$\varepsilon >0$ such that $\omega (t)\in \bar A$ for $0<|t-t_0|<\varepsilon$.
%If we restrict our attention to paths that are nowhere tangent to 
%any of the sets $A_1,\ldots ,A_n$ then  $Z(A_1,\ldots ,A_n)$ is open without
%assuming \rf{star}.  
We claim that any path in $Z(A_1,\ldots ,A_n)$ which is nowhere tangent to
any of the boundaries $\pa A_i$ is an interior point of the set
$Z(A_1,\ldots ,A_n)$.
This can be seen as follows: In addition to the $s_i$'s, 
choose numbers $s_i'\in [0,1]$ such that 
$\tau_i<s_i'<s_i$ and $\omega (s_i')\notin
\bar A_i$.  Now choose $r>0$ smaller than each of the distances from $\omega
(s_i')$ to $\bar A_i$.  It then follows that a path $\bar\omega '$ within 
a distance $r$ from $\bar\omega$ leaves $A_i$ somewhere between $s_i$ 
and $s_i'$ and hence $\bar\omega '\in  Z(A_1,\ldots ,A_n)$ as before.

It is not hard to see that if $\bar\omega\in Z(A_1,\ldots ,A_n) $
is tangent to one of the 
$\pa A_i$'s then $\bar\omega\in \pa Z(A_1,\ldots ,A_n)$, i.e.,
there are paths arbitrarily close to $\bar\omega$ that
are not in $  Z(A_1,\ldots ,A_n)$, see Fig.\ 2.

\begin{figure}[hbt]
  \begin{center}
    \psfrag{x}{$x$}
    \psfrag{y}{$y$}
    \psfrag{z}{$z$}
    \psfrag{A1}{$A_1$}
    \psfrag{A2}{$A_2$}
    \psfrag{A3}{$A_3$}
    \includegraphics[width=12cm]{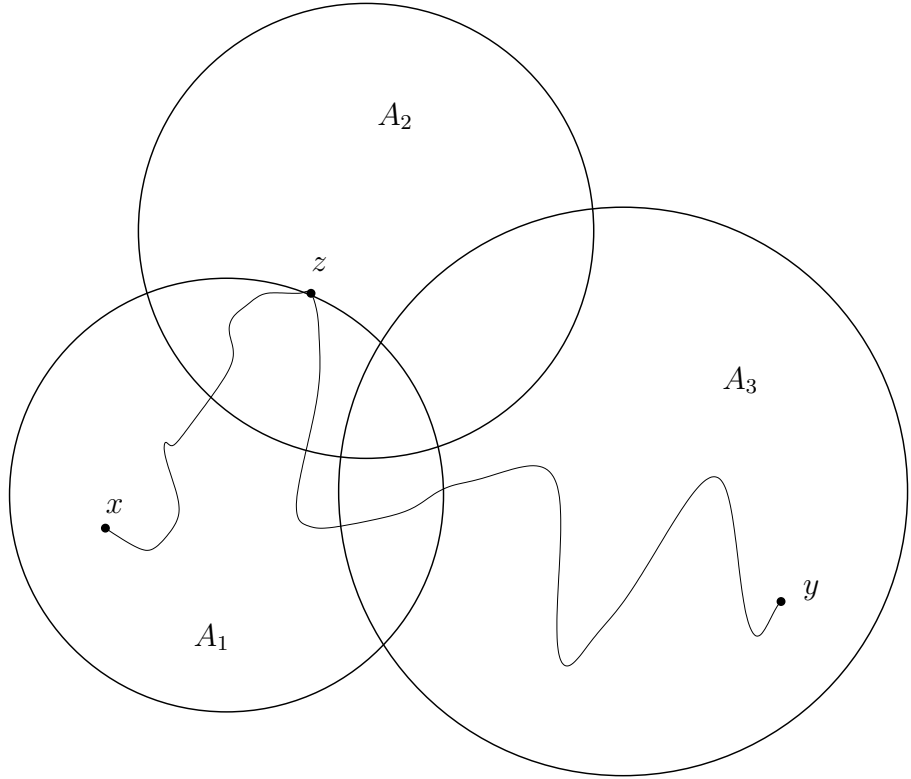}
   
 \caption{A path from $x$ to $y$ which is tangent to $A_1$ at the 
point $z$.  There are paths arbitrarily close to this path which are not in
$Z(A_1,A_2,A_3)$.}
   
    \label{fig2}
  \end{center}
\end{figure}

%But this is not a problem since the boundary of $Z(A_1,\ldots ,A_n)$ is
%contained in a set of measure zero (at least when the $\pa A_i$'s are 
%smooth) as we show in the next section.
We do not have a proof that the sets 
$Z(A_1,\ldots ,A_n)$ are measurable for general open sets 
$A_1,\ldots ,A_n$.
We avoid this problem 
simply by taking the closures of these sets.
We denote the closures by $\bar Z(A_1,\ldots A_n)$.

%however, is not a problem since 
%the Wiener 
%measure of paths that are tangent to a set with smooth boundary is zero.
%This follows from the fact \cite{brownian} that if a Brownian path crosses    
%a plane at time $t=t_0$ then it crosses the plane an infinite number of times
%with probability one during any time interval $[t_0-\varepsilon, t_0+
%\varepsilon]$.  We are therefore free to disregard paths that are 
%tangential.  This will be done in the proof of the following result.

\medskip

\noindent
{\bf Proposition 5.2.} 
The sets  $\bar Z(A_1,\ldots ,A_n) $ where the $A_i$'s 
are open balls generate the
Borel algebra of geometric paths.

\medskip

\noindent
{\it Proof.}  We will show that any open set in $\bar\Omega (x,y)$ can 
be written as a countable union of $\bar Z$-sets.  Given $\bar\omega\in
\bar\Omega (x,y)$ and $\ep >0$ we show that there are open 
balls $A_1,\ldots ,
A_n$ such that $\bar\omega\in \bar Z(A_1,\ldots ,A_n)$ and 
$\bar Z(A_1,\ldots ,A_n)$ is contained in a ball in $\bar\Omega (x,y)$ 
of radius $\ep$ centered on 
$\bar\omega$.  Moreover, the $A_i$'s can be taken to have
rational centers and radii.  It follows then by a standard argument that
the $\bar Z$-sets generate the Borel algebra.

Let $\bar\omega\in U$ where $U\subseteq\bar\Omega (x,y)$ is open.  Choose
a rational number $\ep$ so that $\ep<\oh \bar d(\bar\omega ,\pa U)$.  Let 
$A_1$ be an open 
ball of radius $\ep$ centered at $x$.  If $\bar\omega$ is not 
contained in $A_1$ let $x_1\in \mathbb R^d$ be the point where $\bar\omega$
leaves $A_1$ for the first time, i.e., if $\omega :[0,1]\to \mathbb R^d$
is a parametrization
of $\bar\omega$, then $x_1=\omega (\tau _1)$, where
$$
\tau_1=\sup\{t\in [0,1]:\omega ([0,t])\subseteq A_1\}
$$  
as before.
Take a point $y_1$ with rational 
coordinates such that $|x_1-y_1|<\ep/3$.  Let $A_2$ be a ball of radius 
$\ep$ cenetred at $y_1$.  If $\bar\omega$ stays inside $A_2$ after it leaves
$A_1$ at $x_1$ the construction is finished; otherwise let $x_2$ be the point 
where $\bar\omega$ leaves $A_2$ for the first time after it left $A_1$
at $x_1$ and define $y_2$ and 
$A_3$ in a way analogous to the one used to define $y_1$ and $A_2$.
The construction continues in this way until we obtain a set $A_n$ inside 
which $\bar\omega$ stays after it leaves $A_{n-1}$.  The construction has 
to end after a finite number of steps since any paramterization 
$\omega$ of $\bar\omega$ is a uniformly 
continuous map.   

From the above construction it is clear that 
$\bar\omega\in  
Z(A_1,\ldots ,A_n)$.  Moreover, if $\bar\omega '$ is another
path in $Z(A_1,\ldots ,A_n)$ then $\bar d(\bar\omega ,\bar\omega ')
\leq 2\ep$
because we can choose a parametrization $\omega '$ of $\bar\omega '$ such that
the $\tau_i$'s coincide for $\omega$ and $\omega '$ and hence, for any
$t\in [0,1]$,
$\omega (t)$ and $\omega '(t)$ both belong to the same $A_j$, 
$j=1,\ldots ,n$.  We conclude that 
$Z(A_1,\ldots ,A_n)$ and hence $\bar Z(A_1,\ldots ,A_n)$
is contained in a closed ball in $\bar\Omega (x,y)$ 
of radius $2\ep$ centered on $\bar\omega$.  This ball is contained in $U$
and the proof is complete.
\hfill $\Box$

\medskip

We remark that the proof of the above result can of course be adapted to the 
case where the sets $A_i$ are boxes in $\mathbb R^d$ rather than balls.

\section{Integrating over cylinder sets}

In this section we show that the lattice approximation to the measure of 
the $\bar Z$-sets converges and we 
derive some formulae for the measure of these sets in terms of 
Dirichlet propagators.

Let $A$ be a bounded set in $\mathbb R^d$ with a smooth boundary.
Let $x$ and $y$ be two different points in the interior of $A$.  We recall
that the Dirichlet Green function for
$\oh (-\Delta +m^2)$ with data on $\pa A$, denoted $G_A^D(x,y)$, is given by
the Wiener integral over all paths from $x$ to $y$ that avoid $\pa A$.
This fact is established in, e.g., 
\cite{dirichlet} for the corresponding heat kernel and hence follows for the
propagator by integrating over time.  

In the following discussion the endpoints $x$ and $y$
will be kept fixed and for simplicity we denote the measure $V_{x,y}$ 
by $\mu$.  Accordingly we can write
\begin{equation}
G_A^D(x,y)=\int_{\bar Z(A)}\,d\mu =\mu (\bar Z(A)).\label{chi}
\end{equation}
%where $\chi_A$ is the characteristic function of the set of 
%geometric paths from $x$ 
%to $y$ that hit $A$.  
We are interested in generalizing this formula to
the case of $\bar Z(A_1,\ldots ,A_n)$ with $n>1$ and showing that 
\begin{equation}
\lim_{a\to 0}V_{x,y,a}(\bar Z(A_1,\ldots ,A_n))=\mu (\bar Z(A_1,\ldots ,A_n)).
\label{lconv}
\end{equation}
  In order to minimize
technical complications let us assume that the sets $A_i$ are boxes so their
boundaries are contained in hyperplanes.

Let us consider a family of boxes $A_1,\ldots ,A_n$ in $\mathbb R^d$
with the propery that the intersection of any two different 
boundaries $\pa A_i$ and $\pa A_j$ has codimension 2 or greater, i.e., 
the boundaries never overlap. 
Let us define $O_1$ as the collection of all paths in $\bar\Omega (x,y)$ 
that are somewhere tangent to one of the hyperplanes that make up the 
boundaries of the $A_i$'s.  Let $O_2$ be the collection of all paths in 
$\bar\Omega (x,y)$ that meet one or more of the intersections 
$\pa A_i\cap \pa A_j$, $i\neq j$.  Put $O=O_1\cup O_2$.  It can be checked 
that the set of paths that are somewhere tangent to a given hyperplane 
is a measurable set with measure zero.  It has measure zero since 
the probability that a Wiener path intersects a hyperplane exactly once in 
a time interval is zero, see, e.g., \cite{breimann} Chapter 12.   
The set $O_2$ is easily seen
to be closed and hence measurable.
Its measure is zero since the codimension of the intersections $\pa A_i\cap
\pa A_j$ is greater than $1$.  Thus, $O$ is a measurable set with measure $0$.
The boundary of  $Z(A_1,\ldots ,A_n)$ consists of paths for which 
either two of the 
$\tau_i$'s coincide or the path is tangent to one of the boundaries 
$\pa A_i$.  Hence, 
$\pa  \bar Z(A_1,\ldots ,A_n) \subseteq 
\pa  Z(A_1,\ldots ,A_n)\subseteq O$.  {\it We can therefore conclude from
\cite{billingsley} Theorem 2.1 that the convergence \rf{lconv} takes place
for boxes} and the argument can be extended to the case of
$A_i$'s with piecewise smooth boundaries. 

Let us now turn to the calculation of the measure of the 
$\bar Z$-sets. 
Let $A$ be as before.
Since $G_A^D(x,z)=0$ for $z\in \pa A$ we have
\begin{equation}
\int_{\pa A}\frac{\pa}{\pa n}\left(G_A^D(x,z)G(z,y)\right)\,dS =
\int_{\pa A}\frac{\pa G_A^D}{\pa n}(x,z)G(z,y)\,dS
\end{equation}
where $\frac{\pa}{\pa n}$ is the normal derivative to $\pa A$ with 
respect to $z$.  Let $Y_A$ be the collection of all paths from $x$ to $y$ 
which hit the boundary $\pa A$.  An application of the divergence theorem and
Eq.\ \rf{chi} lead to
\begin{equation}
\mu (Y_A) =\int_{\pa A}\frac{\pa G_A^D}{\pa n}(x,z)G(z,y)\,dS.\label{50}
\end{equation}
More generally, it can be argued that
\begin{equation}
P_A(z)= \frac{\pa G_A^D}{\pa n}(x,z)G(z,y) \label{form1}
\end{equation}
is (up to the constant factor $\mu (Y_A)$) 
the {\it conditional probability density} that a path from $x$ to $y$ which
hits the boundary $\pa A$ 
hits it for the first time at the point $z\in\pa A$, and
$P_A(z)$ is given by an integral over all paths from $x$ to $y$ which 
hit the boundary of $A$ and hit it for the first time in $z$. 

It is convenient to extend the Dirichlet Green functions $G_A^D$ to all of 
$\mathbb R^d$ such that they are $0$ outside $A$.  Let us now consider 
the case $x\in A_1$, $y\in A_2\setminus A_1$, $A_1\cap A_2\neq\emptyset$.
Then the measure of $\bar Z(A_1,A_2)$ is the integral over all paths from $x$
to $y$ which leave $A_1$ for the first time at a point $z\in \pa A_1\cap A_2$
and stay in $A_2$ after they leave $A_1$.  The integral over these paths is 
obtained by analogy with Eq.\ \rf{50} as 
\begin{equation}
\mu (\bar Z(A_1,A_2))=\int_{\pa A_1} 
\frac{\pa G_{A_1}^D}{\pa n}(x,z)G_{A_2}^D(z,y)\,dS
\end{equation}
and
\begin{equation}
 \frac{\pa G_{A_1}^D}{\pa n}(x,z)\frac{G_{A_2}^D(z,y)}{\mu (\bar Z(A_1,A_2)}
\end{equation}
is the conditional probability density that 
a path in $\bar Z(A_1,A_2)$ leaves $A_1$ for the first time in the point $z$.

It is straightforward to generalize the above considerations to the
case of arbitrary $n$, i.e., $\bar Z(A_1,\ldots ,A_n)$.
By the Markov property
of the Brownian paths we have
\begin{equation}
\mu (\bar Z(A_1,\ldots ,A_n)) =\int_{\pa A_1}\ldots \int_{\pa A_{n-1}}
\prod_{i=1}^{n-1}\frac{\pa G_{A_i}^D}{\pa n_i}(z_{i-1},z_i)
G_{A_n}^D(z_{n-1},y)\,dS_1\ldots
dS_{n-1}, \label{bigg}
\end{equation}
where we have set $z_0=x$ and  $\frac{\pa}{\pa n_i}$ denotes the normal 
derivative to $\pa A_i$ with 
respect to $z_i$.  In this integral formula $z_i$ is the point where the
path first leaves $A_i$ after hitting $A_1,\ldots ,A_{i-1}$ in that order.

We note that for $n=1$ 
the convergence \rf{lconv}, i.e., 
the convergence of the lattice approximations to the
Dirichlet propagators is well known for sufficiently nice sets $A$.  
This convergence
can also be proved directly without the use of measure theory.
We also note that all the 
integration formulae above have clear lattice analogues for arbitrary $n$.
%Our main concern in this section has been to establish the convergence
%of these.

\section{Conclusion}

We have in this paper defined integration over geometric paths and
studied natural discretized measures on spaces of such paths.
Two different discretizations were discussed, one with a metric degree of 
freedom and one without.
We have proven the convergence of the discretized measures and thereby
in particular established the convergence of the discrete approximations to
the integrals over paths that one is normally interested in for 
physics applications.
We furthermore introduced, in the case without a metric degree of freedom, 
a natural class of sets of geometric paths which
play the role of cylinder sets and generate the Borel algebra and we have
shown how to calculate the measure of these sets in terms of Dirichlet
propagators.

One, perhaps disappointing but not entirely 
unexpected, outcome of our analysis
is that no technical simplifications are obtained by considering only 
parametrization independent quantities, i.e., by restricting to 
inherently physical degrees of freedom. In
particular, it is hard to get a technical handle on
geometric paths without introducing parametrizations to calculate with as
is usually done in theories with a local gauge invariance.

One of the main motivations for this study was to obtain some insight into
the corresponding problem for random surfaces.  The random surface case
is far more difficult than the one considered in this paper 
 since the measures on parametrized surfaces
which correspond to Wiener measure on paths are not well understood.
Some of the ideas we have discussed here can be carried over to
embedded surfaces but modifications would be needed since points
on a geometric surface cannot be ordered like the points on a geometric path.
For nonimbedded surfaces a new approach is required.  In the absence 
of
imbedding degrees of freedom, points on the surface have to be 
identified in terms of intrinsic geometric degrees of freedom like curvature.
How to do this in a systematic fashion is far from obvious.

\medskip

\noindent
{\bf Acknowledgement.} We are indebted to Institut Mittag Leffler for
hospitality in the spring of 1999 when this work was begun. 
T.J. is also indebted to the Niels Bohr Institute and the 
CERN Theory Division for hospitality.  The work of B.D. is supported in
part by MatPhySto funded by the Danish National Research Foundation.  
This research was partly supported by TMR grant no. HPRN-CT-1999-00161.


\begin{thebibliography}{99}

\bibitem{dewitt}S.W. Hawking and W. Israel (ed.), {\it General relativity -
 An Einstein centennary survey.}  Cambridge University Press, Cambridge 
(1979) 
\bibitem{hawking} S.W. Hawking and G.F.R. Ellis, {\it The large scale
    structure of space time.} Cambridge monographs on
mathematical physics, Cambridge University Press, Cambridge (1973)
\bibitem{book}J. Ambj\o rn, B. Durhuus and T. Jonsson, {\it Quantum geometry
- A statistical field theory approach.}  Cambridge monographs on
mathematical physics, Cambridge University Press, Cambridge (1997)

\bibitem{varadarajan}T. Digernes, V.~S. Varadarajan and S.~R.~S. Varadhan,
Rev. Math. Phys. {\bf 6} (1994) 621

\bibitem{divecchia}L. Brink, P. Di Vecchia and P. Howe, Nucl. Phys.
{\bf 118} (1977) 76

\bibitem{polyakov}A. M. Polyakov, Phys. Lett. B {\bf 103} (1981) 207 

\bibitem{dirichlet} B. Simon, {\it Functional Integration and Quantum Physics},
Academic Press, New York (1979)
\bibitem{billingsley} P. Billingsley, 
{\it Convergence of Probability Measures.} John
  Wiley and Sons, New York (1968)

\bibitem{spitzer} F. Spitzer, {\it Principles of Random Walk.} D. van Nostrand
  Company, Princeton (1964)

\bibitem{breimann}L. Breimann, {\it Probability}, Addison-Wesley, 
Reading (1968)
\end{thebibliography}
\end{document}